\documentstyle[preprint,prl,aps]{revtex}
\begin{document}
 \draft
 \preprint{GTP-98-03}
  \title{Escape rate of a biaxial nanospin system in a magnetic field:
         first- and second-order transitions between quantum and classical regimes}
 \author{Chang Soo Park$^1$, Sahng-Kyoon Yoo$^2$, D. K. Park$^3$ and Dal-Ho Yoon$^4$}
 \address{$^1$Department of Physics, Dankook University Cheonan 330-714, Korea
          \\
          $^2$Department of Physics and Basic Science Research Institute, Seonam
          University, Namwon, 590-711, Korea \\
          $^3$Department of  Physics, Kyungnam University, Masan 631-701, Korea
          \\
          $^4$Department of Physics, Chongju University, Chongju 360-764, Korea}
 \date{January 21, 1999}
 \maketitle

 \begin{abstract}
  We investigate the escape rate of the biaxial nanospin
  particle with a magnetic field applied along the easy axis.
  The model studied here is described by the Hamiltonian ${\cal H}
  = -A S_z^2 - B S_x^2 - H S_z , (A>B>0)$. By reducing this
  Hamiltonian to a particle one, we derive, for the first time, an
  effective particle potential for this model and find an analytical form of the
  phase boundary line between first- and second-order transitions,
  from which a complete phase diagram can be obtained. We also
  derive an analytical form of the crossover temperature as a function of the
  applied field at the phase boundary.
 \end{abstract}

Recently the quantum-classical phase transition of the escape rate \cite{chud} in nanospin system
has been studied intensively. One of the main issues in this subject is to determine whether the
transition is first-order or second-order. In this regards, for the uniaxial spin system such
as high-spin molecular magnet $\rm Mn_{12} Ac$\cite{ziolo}, two models have been investigated
comprehensively : one
with a transverse field \cite{chudgara} and the other with an arbitrarily directed field
\cite{marti}, described by
the Hamiltonians ${\cal H} = -DS_z^2 - H_x S_x $ and ${\cal H} = -DS_z^2 - H_x S_x - H_z S_z $
respectively. In the first case, by using the method of particle
mapping and the Landau theory of phase transition,  Chudnovsky and Garanin have shown that
the transition order changes from first to second when the field parameter $h_x \equiv H_x / (2SD)$
is 0.25. In the case of model with arbitrarily directed field Garanin et al. have
obtained the phase boundary line $h_{xc} = h_x (h_z )$ to show the whole phase diagram
in which $h_{xc} (0) = 0.25$ in the unbiased case  and $h_{xc} \sim (1-h_z )^{3/2}$ in the
strongly-biased limit.

For biaxial spin system such as iron cluster $\rm Fe_8$ \cite{ohm} Liang et al. considered a
model without an applied field, ${\cal H} = K(S_z^2 + \lambda S_y^2 ), (0< \lambda <1)$ \cite{liang}.
Using the coherent spin
state representation they have shown that the coordinate dependent effective mass leads
to the first-order transition and the change between the first- and second-order
transitions occurs at the value $\lambda = 0.5$ . The biaxial spin model with transverse field,
${\cal H} = K(S_z^2 + \lambda S_y^2 ) - H_y S_y $, has also been investigated
by Lee et al. who demonstrated that various types of
combinations of the first- and second-order transitions are
possible depending on $\lambda$ and $H_y /K\lambda S$ \cite{lee}.

In this paper we study the phase transition of the escape rate of the biaxial spin system
with a longitudinal field. We will first derive an effective particle potential by mapping the
spin Hamiltonian onto particle one, which is the first derivation for the present model.
Then, with the help of the recently developed method
for the criterion of the transition order \cite{gorv,dkpark}, we find an analytical form of the phase
boundary line from which a complete phase diagram for the order of the phase transition is obtained.

Consider a nanospin particle with an applied field $H$ along the easy axis. If the
spin particle is a biaxial spin system with $XOZ$ easy plane  anisotropy and the easy
$Z$-axis in the $XZ$-plane the Hamiltonian can be described by
\begin{equation}
{\cal H} = -AS_z^2 - BS_x^2 - HS_z
\end{equation}
where the anisotropy constants satisfy $A>B>0$. Our model is equivalent to
${\cal H} = K(S_z^2 + \lambda S_y^2 ) - HS_x , (\lambda < 1) $ if we set
$A = K, B = (1-\lambda)K $. In the following, for convenience,
we introduce dimensionless anisotropy parameter $b \equiv B/A$ and field
parameter $\alpha \equiv H/SA ,  (0< \alpha <2)$ where $S$ is the spin number. For iron cluster
${\rm Fe}_8$ in Ref.5, $S = 10, A = 0.316$ K, and $B = 0.092$ K.
We can reduce this spin problem to a
particle moving in a potential \cite{zaslav}. The equivalent Schr\"{o}dinger-like
equation is derived as
\begin{equation}
- \frac{1}{2m} \frac{d^2 \Psi} {dx^2 } + V(x)\Psi = E\Psi,
\end{equation}
where $m = 1/2A$,
\begin{equation}
\Psi (x) = \left(\frac{{\rm cn}x}{{\rm dn}x}
\right)^S
\exp \left[ \frac{{\alpha}S}{2\sqrt{1-b}}{\tanh}^{-1} \left(
\sqrt{1-b}{\rm sn}x \right) \right] \Phi(x)
\end{equation}
with
\begin{displaymath}
\Phi(x) = \sum_{\sigma = -S}^{S} \frac{C_{\sigma}}{\sqrt{(S -
\sigma )! (S + \sigma )!}} \left( \frac{{\rm sn}x + 1}{{\rm cn}x}
\right)^\sigma
\end{displaymath}
is the particle wave function, and $V(x)$ is the effective particle potential given by
\begin{equation}
\frac{V(x)}{A} = \frac{\alpha^2 S^2 {\rm cn}^2 x - 2 \alpha bS(2S + 1){\rm sn}x - 4bS(S+1)}
{4 {\rm dn}^2 x}
\end{equation}
in which ${\rm sn}x, {\rm cn}x,$ and $ {\rm dn}x$ are the Jacobian Elliptic functions with modulus $k^2 = 1-b$.
This potential is shown in Fig.1. The local minimum represents
a metastable state of the spin system described by the Hamiltonian (1).
The potential preserves the symmetry of a rotation from $-Z$ to $+Z$-axis.
Thus, the escape from
the local minimum $-x_m$ to the global minimum $x_m$ corresponds to the
inversion of the spin magnetization vector.
For large spin such as $ S(S+1) \sim {\tilde{S}}^2 \equiv (S+ 1/2)^2 $ it has a maximum at point
$x_0 = {\rm sn}^{-1} [-\alpha /2(1-b)]$.
For a given value of $b$ the barrier height decreases as $\alpha$
increases and vanishes at $\alpha = 2(1-b)$ which corresponds to the coercive field.

The type of the phase transition is determined
by the behavior of the Euclidean time oscillation period $\tau(E)$, where
$E$ is the energy, near the bottom of the Euclidean potential
which corresponds to the top of the potential barrier \cite{chud}. When $\tau(E)$
decreases monotonically as $E$ approaches the barrier top the
transition from quantum to classical regime is second-order. If,
however, $\tau(E)$ is not a monotonous function of energy a
first-order quantum-classical transition takes place. One can
also argue that the condition for the first-order transition can
be obtained by looking at the behavior of the oscillation period
in Euclidean time as a function of oscillation amplitude near the
barrier top. In this case the first-order transition appears when
the amplitude dependent period $\tau(a)$, where $a$ is the
amplitude, is smaller than the zero amplitude period $\tau(0)$ which
corresponds to the solution  near the position of the sphaleron solution \cite{gorv}. The
important feature of these approaches relies on the shape of the
potential near the top of the barrier. Thus, parameterizing the
amplitude as the coefficient of the perturbation expansion a
sufficient condition for the first-order transition can be derived
\cite{gorv,dkpark}. Below, we show that the latter approach leads to an
analytical form of phase boundary between first- and second-order
transitions for the present model.

Expanding the potential $V(x)$ near $x=x_0$ up to fourth-order we
obtain
\begin{equation}
V(z+x_0 ) \cong a_1 (x_0 ) z^2 + a_2 (x_0 ) z^3 +a_3 (x_0 ) z^4,
\end{equation}
where $z = x-x_0 , a_1 (x_0 ) = V''(x_0 )/2 (<0), a_2 (x_0 ) =
V'''(x_0 )/6 (>0)$, and $a_3 (x_0 )=V''''(x_0 )/24$. Following
references 8 and 9 the criterion of the first-order
transition can be obtained from the condition $\tau (a) - \tau (0) <
0$, which is expressed by
\begin{equation}
-\frac{15}{4} \frac{a_2^2 (x_0 )}{a_1 (x_0 )} + 3a_3 (x_0 ) < 0 .
\end{equation}
By equating both sides we can find an equation of the phase
boundary line. For large spin number this is obtained to be
\begin{equation}
\alpha_c = 2(1-b_c ) \sqrt{\frac{1-2b_c }{1+b_c }},
\end{equation}
where $\alpha_c$ and $b_c$ are the critical values of the field and
anisotropy parameters on the phase boundary, respectively.
In Fig.2 we have plotted $b_c$ as a function of $\delta_c \equiv 1 -
\alpha_c /2$. This picture displays a complete phase diagram
in the whole range of the applied field. From the condition in Eq.(6)
the first-order exists below the line. From the picture we immediately
see  $b_c = 0.5$ in the unbiased case $\delta_c = 1$.
This is the same as the previously obtained result \cite{liang} and confirms that the present
approach is correct. In the strongly biased case, $\delta_c \rightarrow 0$ and $b_c \rightarrow 0$
since $\alpha_c \rightarrow 2$. Thus, in this limit,
by expanding the Eq.(7) for small $b_c$ we find the linear behavior $b_c \sim 0.4\delta_c$.

For the critical temperature at the boundary between the first- and
second-order transitions we use the formula $2 \pi T_c = \sqrt{- V''(x_0
)/m}$. For the present potential  this becomes
\begin{equation}
\frac{T_c }{\tilde{S} A} = \frac{2\sqrt{3} b_c
}{\pi}\sqrt{\frac{1-b_c }{1 + b_c }} .
\end{equation}
In Fig.3 we have plotted  $T_c / {\tilde{S}}A$ vs. $\delta_c $ graph,
where the relation in Eq.(7) has been
used. In the unbiased case Eq.(8)  gives $T_c /{\tilde{S}}A = 0.318$ which coincides with
the value calculated from the Eq.(13) in Ref.6. In the strongly biased limit it
represents linear behavior $T_c /{\tilde{S}}A \sim 0.442\delta_c $.

Very recently the same spin system has also been considered \cite{cond},
but with a slightly different model :
${\cal H} = -D S_z^2 + BS_x^2 - HS_z, (D,B>0)$. Comparing this with our
model we realize $D = \lambda A$. By using a perturbative approach with
respect to $b \equiv B/D$ they obtained a phase
diagram in the whole range of field from which the linear dependences
of $b_c$ and $T_c / SD$ on
$\delta_c$ in the strongly biased case can be found. Since the
perturbation parameter is $b$ the validity of this approach is
limited in the range of small values of $b$ which requires the
strong bias limit to change the order of transition.
On the other hand,
there is no restriction to $b$ in our approach. We believe that
the present approach is more rigorous and the results are improved.

Finally, we comment on the experimental observation of the results.
Using the anisotropy constants given in Ref.5  we have $b_c =
0.29$, and thus $\alpha_c = 0.81$ from Eq.(7) for which the critical field
is estimated to be 1.9 ${\rm T}$, and the coercive field is 3.4 ${\rm T}$.
From Eq.(8) the transition temperature on the phase
boundary can also be calculated, and we find $T_c = 0.79$ ${\rm K}$.

To summarize we have investigated the phase transition of the
escape rate from metastable state in nanospin system with a
magnetic field applied along the easy axis. By using the particle
mapping we derived an effective particle potential.
We obtained an analytical form for the equation of the phase boundary line
between the first- and second-order transitions and thus a
complete phase diagram. In the strongly biased case we found a
linear dependence of $b_c$, the dimensionless anisotropy parameter
on the applied field. We also obtained a diagram for the
crossover temperature as a function of the applied field. It also
shows a linear relation. The results obtained here can be used as
a guide for the experimental observation.

 \newpage
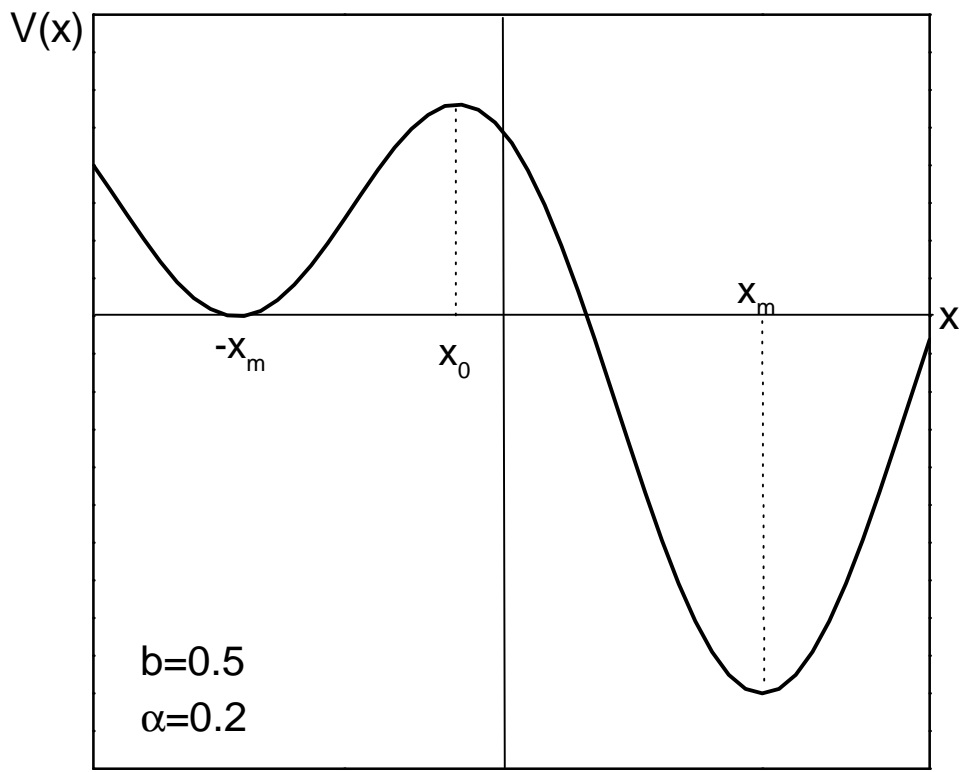
 \begin{figure}
 \caption{Effective particle potential}
 \label{fig1}
 \end{figure}

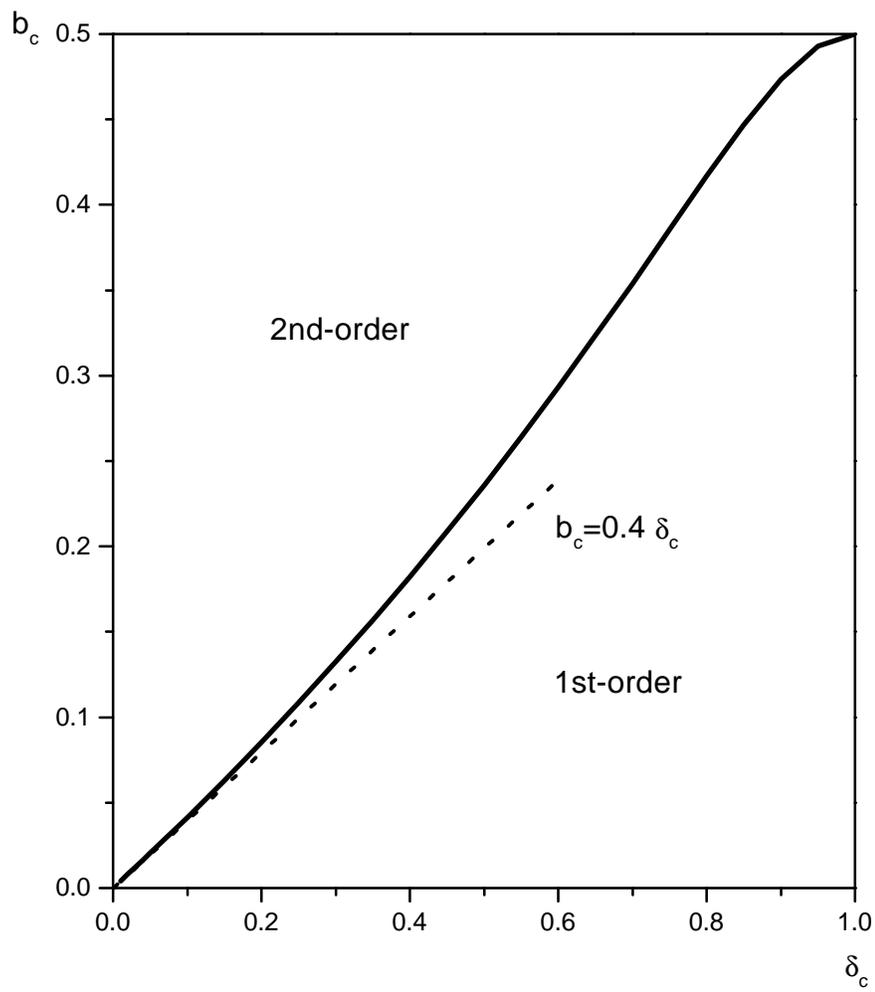
\begin{figure}
\caption{The phase boundary line  $b_c$ vs. $\delta_c$.
Solid line: anaytical result for large
spin number, dashed line: straight line in the strongly biased case.}
\label{fig2}
\end{figure}

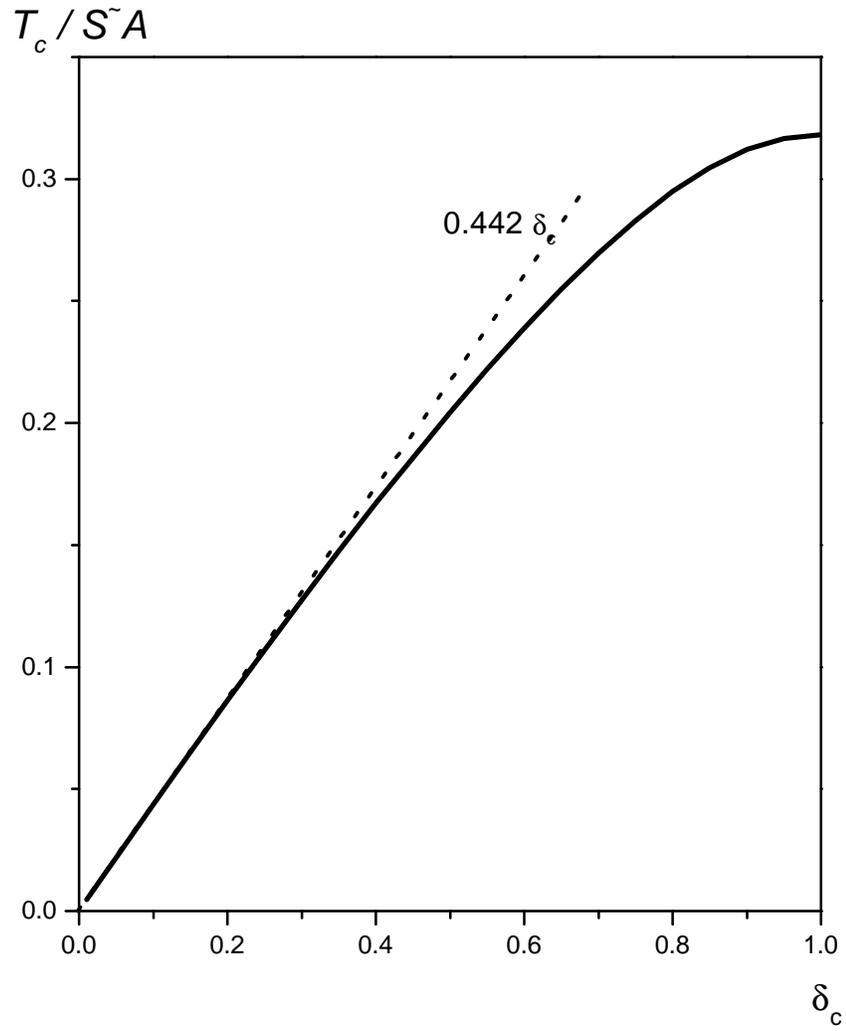
\begin{figure}
\caption{Critical temperature $T_c / \tilde{S} A$ as a function of $\delta_c$.
Respective lines represent the same as Figure 2.}
\label{fig3}
\end{figure}

\end{document}